\documentclass[aps,twocolumn,showpacs,floatfix]{revtex4}

\usepackage{amsfonts}
\usepackage{amsmath}
\usepackage{amssymb}
\usepackage{graphicx}
\usepackage{color}
\usepackage{verbatim}
\usepackage{lipsum}
\usepackage{dsfont}
\usepackage{upgreek}

\usepackage{hyperref}
\usepackage{makecell}

\begin{document}
\title{Dynamic $^{14}\rm N$ nuclear spin polarization in nitrogen--vacancy centers in diamond}
\author{Laima Busaite}
\email{laima.busaite@lu.lv}
\author{Reinis Lazda}
\email{reinis.lazda@lu.lv}
\author{Andris Berzins}
\author{Marcis Auzinsh}
\author{Ruvin Ferber}
\author{Florian Gahbauer}

\affiliation{Laser Centre, University of Latvia, Rainis Boulevard 19, LV-1586 Riga, Latvia}



\begin{abstract}

We studied  
the dynamic nuclear spin polarization of nitrogen in negatively charged nitrogen-vacancy (NV) centers in diamond both experimentally and theoretically over a wide range of magnetic fields from 0 to 1100 G covering both the excited-state level anti-crossing and the ground-state level anti-crossing magnetic field regions.
Special attention was paid to the less studied ground-state level anti-crossing region. The nuclear spin polarization was inferred from measurements of the optically detected magnetic resonance signal.
These measurements show that a very large (up to $96 \pm 2\%$) nuclear spin polarization of nitrogen can be achieved over a very broad range of magnetic field starting from around 400 G up to magnetic field values substantially exceeding the ground-state level anti-crossing at 1024 G.
We measured the influence of angular deviations of the magnetic field from the NV axis on the nuclear spin polarization efficiency and found that, in the vicinity of the ground-state level anti-crossing, the nuclear spin polarization is more sensitive to this angle than in the vicinity of the excited-state level anti-crossing. Indeed, an angle as small as a tenth of a degree of arc can destroy almost completely the spin polarization of a nitrogen nucleus. In addition, we investigated theoretically the influence of strain and optical excitation power on the nuclear spin polarization.
\end{abstract}

\maketitle

\section{Introduction}

Nitrogen-vacancy (NV) centers in diamonds, which consist of a vacancy in the diamond lattice adjacent to a substitutional nitrogen atom, exhibit many characteristics that make them suitable for quantum metrology and quantum information applications~\cite{Popkin2016,Ladd2010,Heshami2016}. In particular, NV centers have triplet ground and excited states, which undergo Zeeman splitting in a magnetic field and exhibit de-excitation dynamics that allow the ground state to be polarized to the spin-zero state through optical pumping. Furthermore, the electron spin can remain polarized for times that are long enough~\cite{Balasubramanian2009}, which allows to measure, for example, the local magnetic field~\cite{Zheng2017,Taylor2008,Fescenko2019a,Wolf2015,Fescenko2019}. However, these measurements become significantly more sensitive if the electron spin polarization can be transferred to the even longer-lived nuclear spin polarization. The transfer of the optically created polarization of the electron spin of the NV center to a nearby nuclear spin, such as the nuclear spin of the NV center's nitrogen atom, is called dynamic nuclear polarization (DNP). For example, DNP in diamond crystals was suggested as an alternative method to achieve very high nuclear spin polarization at moderate magnetic field values and room temperature in order to enhance nuclear magnetic resonance (NMR) signals~\cite{Pagliero2018}, which has already been demonstrated using NV centers as probes~\cite{Pham2016,DeVience2015}. One approach is to enhance the polarization of the nuclear spins of $^13$C in the vicinity of the NV centers~\cite{King2015}. Even more useful would be the ability to achieve DNP in nuclei on the surface of the diamond or in analytes on the surface~\cite{Fernandez-Acebal2018}. Using NV centers as nuclear magnetic resonance probes~\cite{Wood2016,Smits2019} would open up new perspectives reaching down to the cellular or even molecular scale for NMR, which is one of the workhorse techniques of analytical chemistry. Better understanding of the mechanisms involved in DNP could be also helpful in other applications, such as spin preparation for quantum computing.

DNP  has been applied successfully to the nitrogen atom of an NV center in the vicinity of the excited-state level anticrossing (ESLAC) around 512~G for single $^{14}\mathrm{N}$ and $^{15}\mathrm{N}$ spins~\cite{Jacques2009,Ivady2015}, ensembles of $^{14}\mathrm{N}$ spins, and even for proximal $^{13}\mathrm{C}$ spins~\cite{Fischer2013a, Poggiali2017}. The hyperfine interaction in the magnetic field near the electron-spin anticrossing points creates hyperfine states that can couple the spin from the electron to a nucleus or from one nucleus to another.
Experimental signals that measure nuclear spin polarization have been described successfully with models based on rate equations~\cite{Jacques2009} or the Liouville equation for the density matrix, combined with the Lindblad operator~\cite{Lindblad1976}, which can describe transitions between the different spin projection states~\cite{Fischer2013a,Poggiali2017,Ivady2015}.
Another level anticrossing case, namely the ground-state level anticrossing (GSLAC), occurs at 1024~G. Models developed in~\cite{Ivady2015} have predicted that, in the case of $^{15}\mathrm{N}$, the nuclear spin polarization should fall as the magnetic field is increased from the ESLAC to the GSLAC, with a narrower peak at the GSLAC.
Nuclear spin polarization at the ESLAC has been shown to be very sensitive to any angular deviations of the magnetic field from the NV axis~\cite{Jacques2009}, which is important to take into account in any practical applications.
The magnetic field angular deviation influence on nuclear spin polarization has not been studied before at the GSLAC magnetic field region. 

The purpose of the current work is to measure DNP of $^{14}\mathrm{N}$ over a wide range of magnetic field values that includes both the ESLAC and GSLAC, and to compare the measurements with calculations based on a theoretical model.
We measured optically detected magnetic resonance (ODMR) signals over the magnetic field range of 0--1100~G and from these signals calculated the nuclear polarization. As a result, we could infer information about hyperfine transitions. 
We used an experimental setup similar to our previous works~\cite{Auzinsh2019,Lazda2020} with an improved resolution that allowed us in most cases to resolve individual hyperfine components. We calculated the expected positions of the hyperfine transition peaks at a fixed magnetic field. Then we fitted the experimentally measured and sometimes only partially resolved ODMR spectra with Lorentz-shaped contours, and so obtained the relative populations of the basis states. This fitting procedure allowed us to determine the DNP of $^{14}\mathrm{N}$ in NV centers following a similar procedure as in our previous study~\cite{Auzinsh2019}.

We performed numerical calculations of DNP around the ESLAC and GSLAC using a model based on the Liouville equations for the density matrix with a Lindblad operator.
We examined theoretically as well as experimentally
the dependence of DNP and consequently the dependence of the measured ODMR signals on the possible small misalignment of the external magnetic field along the axis of NV center.
The results show that the polarization right at the GSLAC is even more sensitive to the magnetic field alignment than it is in the case of the ESLAC~\cite{Jacques2009}. In fact, a change in the angle between the magnetic field direction and the axis of the NV center as small as $0.1^\circ$ can transform a DNP enhancement into a complete depolarization. Transverse strain~\cite{Doherty2012, Udvarhelyi2018, Barfuss2019, Kehayias2019, Barry2020} in a diamond crystal can also cause nuclear spin depolarization at the GSLAC.

\section{Method}

\begin{figure}[!t] 
  \begin{center}
    \includegraphics[width=0.45\textwidth]{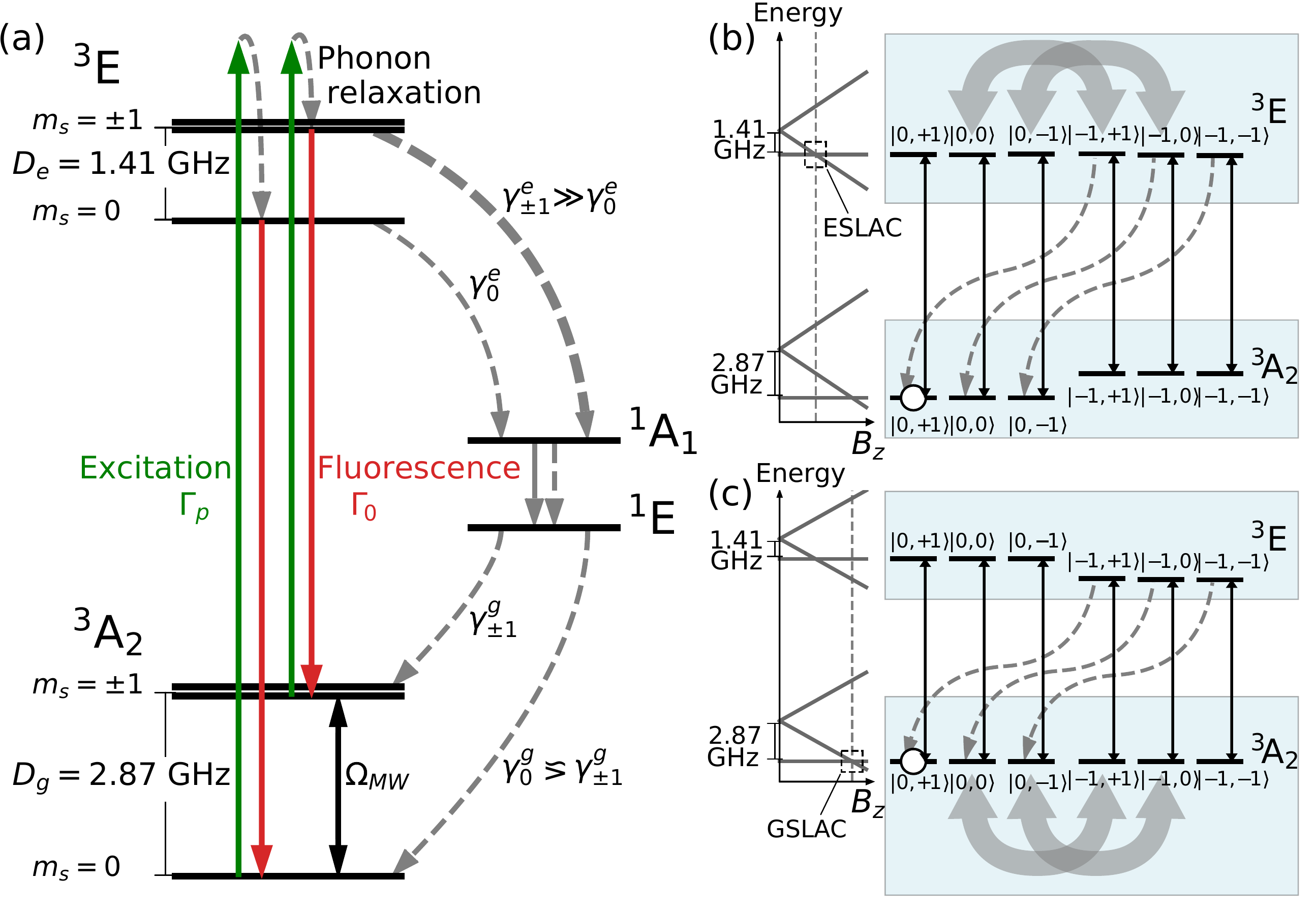}
  \end{center}
  \caption{Left: (a) the NV center's energy-level structure. Right: The nuclear spin polarization process at the (b) ESLAC and (c) GSLAC. Dashed lines indicate nonradiative transitions and straight lines represent the optical transition, which conserve nuclear spin.}
  \label{fig_levels}
\end{figure}

The NV center's ground and excited states are triplet states [Fig.~\ref{fig_levels}(a)] with an electronic spin $S=1$ that interacts with the nuclear spin $I=1$ of the substitutional $^{14}\mathrm{N}$.
At room temperature the NV-center system can be described by the same form of the Hamiltonian for both the ground and the excited states~\cite{Doherty2013}. The ground state Hamiltonian in the presence of an external magnetic field $\mathbf{B}$ can be written as: 
\begin{equation}
\begin{split}
 \hat{H_g} &= D_g \hat S_z^2 + \gamma_e \mathbf{B} \cdot \mathbf{\hat S} + Q \hat I_z^2 + \mathbf{\hat S} \cdot \bar{A}_g \cdot \mathbf{\hat I} - \gamma_\text{14N} \mathbf{B} \cdot \mathbf{\hat I} \\
 &  + N^g_{x}\left(\hat S_x \hat S_z + \hat S_z \hat S_x\right), 
 \end{split}
 \label{eq_ham_ground}
\end{equation}
where $\mathbf{\hat S}$, $\hat S_z$ and $\mathbf{\hat I}$, $\hat I_z$ are the electronic and nuclear spin and their $z$-projection operators. The first term describes the zero-field splitting of the electron-spin sublevels $m_S=0$ and $m_S=\pm 1$ by $D_g = 2.87~\mathrm{GHz}$~\cite{Doherty2013}. The second term is the electronic spin interaction with the magnetic field vector $\mathbf{B} = \left(B\sin\theta,~0,~B\cos\theta \right)$, where $\theta$ is the angle between magnetic field direction and the NV center's principal axis, $\gamma_e=2.8025~\mathrm{MHz/G}$ is the gyromagnetic ratio of the spin of the NV center. The third term is the quadrupole splitting of the $^{14}\mathrm{N}$ nucleus, $Q = -4.96~\mathrm{MHz}$. The fourth term is the hyperfine interaction of the electron spin and the nuclear spin of the NV center, where $\bar{A}_g$ is the ground-state hyperfine interaction tensor. The fifth term of Eq.~\eqref{eq_ham_ground} describes the nuclear-spin interaction with the magnetic field ($\gamma_{14N}=0.3077~\mathrm{kHz/G}$ is the gyromagnetic ratio of the $^{14}\mathrm{N}$ nuclear spin)~\cite{Wood2016}.
The influence of the strain is included in the last term of the Hamiltonian, where $N^g_{x}$ is the ground-state transverse strain component that couples the $\hat S_x$ and $\hat S_z$ projections of the electronic spin operator~\cite{Kehayias2019, Barfuss2019}.
At the GSLAC, where the strain influence is investigated, the ground-state sublevels $m_S=0$ and $m_S=-1$ are close enough to be influenced by the strain coupling term $N^g_{x}$. We neglect other strain coupling terms in this work, because at the GSLAC other levels are too far from each other to be influenced by strain coupling.

The hyperfine interaction of the $^{14}\mathrm{N}$ nuclear spin with the electron-spin angular momentum of the NV center can be written in the form
\begin{equation}
    \hat H^g_\text{hfs} = \mathbf{\hat S} \cdot \bar{A}_g \cdot \mathbf{\hat I} = A^g_\parallel \hat I_z \hat S_z + A^g_\perp\left( \hat S_x \hat I_x + \hat S_y \hat I_y\right),
    \label{eq_hfs}
\end{equation}
where the hyperfine-interaction parameters are $A^g_\parallel = - 2.16~\mathrm{MHz}$ and 
$A^g_\perp = -2.70~\mathrm{MHz}$ \cite{Smeltzer2009, Felton2009}.

The excited-state Hamiltonian $H_e$ has the same form as Eq.~\eqref{eq_ham_ground}, with a zero-field splitting of $D_e = 1.42~\mathrm{GHz}$~\cite{Doherty2013}.
The excited-state hyperfine interaction Hamiltonian $\hat H^e_\text{hfs}$ is similar to Eq.~\eqref{eq_hfs}, but with a hyperfine interaction tensor $\bar{A}_e$ 
with $A^e_\parallel = -40~\mathrm{MHz}$ and 
$A^e_\perp = -23~\mathrm{MHz}$~\cite{Poggiali2017}.

At a magnetic field value that corresponds to the level anticrossing (at $\approx\! 512~\mathrm{G}$ in the excited state and $\approx\! 1024~\mathrm{G}$ in the ground state), the magnetic sublevels with $m_S = -1$ approach the magnetic sublevels with $m_S = 0$, and through the hyperfine interaction the state $\vert m_S=0, m_I=0\rangle$ mixes with $\vert m_S=-1, m_I=+1\rangle$, and the state $\vert m_S=0, m_I=-1\rangle$ mixes with $\vert m_S=-1, m_I=0\rangle$.
The mixing of the states as well as the continuous cycle of excitation and nonradiative decay polarize the nuclear spin into the unmixed $\vert m_S=0, m_I=1\rangle$ state [Fig.~\ref{fig_levels}(b)].
The details of the polarization process at the ESLAC and GSLAC are similar and are shown in Fig.~\ref{fig_levels}(b) and (c).
Let us look in more detail into the nuclear spin polarization process in the case of the ground state. If at first we neglect the nuclear spin of $^{14}\mathrm{N}$ and examine Fig.~\ref{fig_levels}(a), we can explain how the NV center is polarized by being pumped into the ground-state electron-spin magnetic sublevel $m_S = 0$.
After light absorption and rapid relaxation in the phonon band, the excited-state magnetic sublevels either can decay back to the ground state with equal rates $\Gamma_0$ and radiate light with a wavelength in the red part of the electromagnetic spectrum or undergo nonradiative transitions to the singlet state $^{1}A_1$ [Fig.~\ref{fig_levels}(a)]. 
The nonradiative transition rate from the excited-state $^3E$ sublevels $m_S = \pm 1$ to the singlet state $^1A_1$ is several times higher than from the excited state $m_S = 0$ to the singlet state. In addition, the nonradiative transition rate from the singlet $^{1}E$ state to the triplet ground-state $^3A_2$ sublevel  $m_S = 0$ is slightly larger than to ground-state $^3A_2$ sublevels $m_S = \pm 1$~\cite{Dumeige2013,Poggiali2017}.
The differences between the nonradiative transition rates imply that after several excitation-relaxation cycles the population of the NV center's triplet ground state will be transferred to the ground-state magnetic sublevel $m_S = 0$, and the electron spin angular momentum will be polarized~\cite{Doherty2013}.
If we now consider the nuclear spin $I$ as well, we see that the optical electron-spin polarization process populates three hyperfine states $\vert m_S=0, m_I=0, \pm 1\rangle$ and depletes three states $\vert m_S=-1, m_I=0, \pm 1\rangle$. But at the ESLAC and GSLAC, the states $\vert m_S=0, m_I= 0, -1\rangle$ are mixed with their counterparts from the $m_S = -1$ manifold. For this mixing to occur the sum of the electron and nuclear angular momentum projections for the mixing partners must be equal~\cite{Budker2004}. As can be seen in Fig.~\ref{fig_levels}(c), from the $m_S = 0$ manifold only the hyperfine level $\vert m_S=0, m_I +1\rangle$ remains unmixed and consequently preserves its population. Therefore, it is the level mixing in the vicinity of the GSLAC that allows the nuclear spin of $^{14}\mathrm{N}$ to be polarized into the $m_I = +1$ state.

As the ground-state hyperfine interaction is much weaker than the excited-state hyperfine interaction~\cite{Poggiali2017, Smeltzer2009, Felton2009}, the polarization of the $\vert m_S=0, m_I=1\rangle$ state at the GSLAC is expected to be influenced more by external processes, such as a transverse magnetic field $B_x$ (slight misalignment of the magnetic field from NV-axis) or strain in the diamond crystal.

\section{Numerical model}

To calculate the nuclear spin polarization over a wide range of magnetic field values, we used a numerical model that takes into account the ground state $^3A_2$ and the excited state $^3E$ with the nuclear spin $I=1$ of the $^{14}\mathrm{N}$. The system is described using a density operator $\rho$ that accounts for 21 states: 9 in the ground state, 9 in the excited state, and 3 the in singlet state. Thus, in the decay channel from the excited triplet state $^3\mathrm E$ to the ground triplet state $^3\mathrm A_2$ via the two singlet states $^1\mathrm A_1$ and $^1\mathrm E$ in a cascade-type transition, two singlet states are substituted in the model by one singlet state with three nuclear spin components. 

The nuclear spin polarization is calculated from a steady-state solution of the Lindblad equation~\cite{Lindblad1976}:
\begin{equation}
\frac{\partial\rho}{\partial t} = -\frac{i}{\hbar} \left[\hat{H},\rho\right] + \hat L\rho = 0,
\end{equation}
where $\hat H$ is the Hamiltonian of the ground and excited states, Eq.~\eqref{eq_ham_ground}, and $\hat L$ is the Lindblad superoperator~\cite{Ticozzi2009}, which is used to describe the depopulation and decoherence processes of the NV center electron spin and $^{14}\mathrm{N}$ nuclear spin:
\begin{equation}
    \hat L\rho = \sum_{k} \hat\Gamma_{k} \left( L_{k} \rho L_{k}^\dagger -\frac{1}{2}\left\lbrace L_{k}^\dagger L_{k},\rho \right\rbrace\right).
\end{equation}
In this equation $\hat\Gamma_k$ are the decay rates associated with the relaxation processes, and the operators $L_k$ describe the depopulation and dephasing or decay of the diagonal and off-diagonal elements of the density matrix $\rho$ (processes characterized by time constants $T_1$ and $T_2$). The operator $L_k =\vert i\rangle\langle j\vert$ corresponds to transitions between states $\vert i\rangle$ and $\vert j\rangle$, which describe the depopulation.
Transitions described by this operator include the transition between the triplet ground and excited states (fluorescence rate $\Gamma$ and pumping rate $\Gamma_p$), the intersystem crossing (non-radiative transition) between the triplet excited states and the singlet states $\gamma^e_0$ and $\gamma^e_{\pm 1}$ as well as from the singlet states to the triplet ground states $\gamma^g_0$ and $\gamma^g_{\pm 1}$. This operator also describes $T_1$-related processes for electronic and nuclear spin (population transfer between magnetic sublevels in the ground and excited states). The operator $L_k = \vert i\rangle\langle i\vert-\vert j\rangle\langle j\vert$ corresponds to the decoherence of the states $\vert i\rangle$ and $\vert j\rangle$. This operator describes $T_2$-related processes in the ground and excited states for electronic and nuclear spin.
The pumping rate $\Gamma_p$ is included in the Lindblad operator as a transition rate between the ground and excited states. The pumping rate was set to $\Gamma_p=5$~MHz for all of the theoretical calculations, except for the power dependence.

\begin{table}[t]
\setlength{\tabcolsep}{6pt}
\begin{tabular}{l  l} 
 \hline
 \hline
 Transition  & Rate \\  
 \hline
 Spin conserving transitions $\Gamma$ & 66 MHz \\ 
 Intersystem crossing $\gamma^e_{\pm 1}/\gamma^e_0$ & 20 \\
 Intersystem crossing $\gamma^g_0/\gamma^g_{\pm 1}$ &  1 \\
 Ground state $T_1$, NV electron spin & 10 ms\\
 Ground state $T_2$, NV electron spin & 100 $\rm\upmu s$\\
 Ground state $T_1$, $^{14}$N nuclear spin & 10 s\\
 Ground state $T_2$, $^{14}$N nuclear spin & 10 $\rm\upmu s$\\
 Excited state $T_1$, NV electron spin & 1 ms\\
 Excited state $T_2$, NV electron spin & 10 ns\\
 Excited state $T_1$, $^{14}$N nuclear spin & 100 ms\\
 Excited state $T_2$, $^{14}$N nuclear spin & 1 ms\\[1ex] 
 \hline\hline
\end{tabular}
\caption{Transition rates and time constants used in the calculation.}
\label{table_decay_rates}
\end{table}

In the calculations we used the values for the transition rates given in~\cite{Dumeige2013,Tetienne2012} and for the longitudinal and transverse relaxation time constants $T_1$ and $T_2$ given in~\cite{Scharfenberger2014, Fischer2013}, which are typical for the type of sample used in the experiment (see Table~\ref{table_decay_rates}).
The electron-spin polarization depends on the nonradiative transitions rates (the ratio of the rates) to and from the singlet state, which characterize the population transfer from $m_s=\pm1$ to $m_s=0$ and from $m_s=0$ to $m_s=\pm 1$. In the model we set the ratio of the transition rates from the singlet state to the triplet ground state $m_s=0$ and from the singlet state to the triplet ground state $m_s=\pm 1$ to $\gamma^g_0/\gamma^g_{\pm 1} = 1$, and adjusted the ratio of the transition rates from the excited state $m_s = \pm 1$ to the singlet state and from the excited state $m_s = 0$ to the singlet state to match the typical electronic spin polarization that has been observed in experiments~\cite{Harrison2006}; 
it is set to $\gamma^e_{\pm 1}/\gamma^e_0=20$, which is close to a value used in~\cite{Fischer2013a}.

From the steady state solution of the density matrix, we calculated the nuclear spin polarization as 
\begin{equation}
P_{th} = \frac{\rho_{01}-\rho_{0-1}}
{\rho_{01}+\rho_{00}+\rho_{0-1}},\label{eq_theor_polarization}
\end{equation}
where $\rho_{m_S m_I} = \langle m_S, m_I \vert \rho_{SS}\vert m_S, m_I\rangle$ is the population of the ground-state basis state $\vert m_S, m_I \rangle$.

\section{Experimental setup}

We measured ODMR signals from ensembles of NV centers in a diamond sample at room temperature (around $20~^\circ$C).
The sample used in the experiment was a diamond crystal obtained through chemical vapor deposition (CVD) with a (100) surface cut.
The diamond sample had been created with a nitrogen $^{14}$N concentration of about 5--20~ppm. The crystal had been irradiated with electrons ($3\cdot10^{17}$cm$^{-2}$ $e^-$ irradiation dosage at 2 MeV) and annealed afterwards in a two-step process, first at $800~^\circ$C and then $1100~^\circ$C, creating NV centers in the bulk of the diamond.

\begin{figure}[t]
  \begin{center}
    \includegraphics[width=0.45\textwidth]{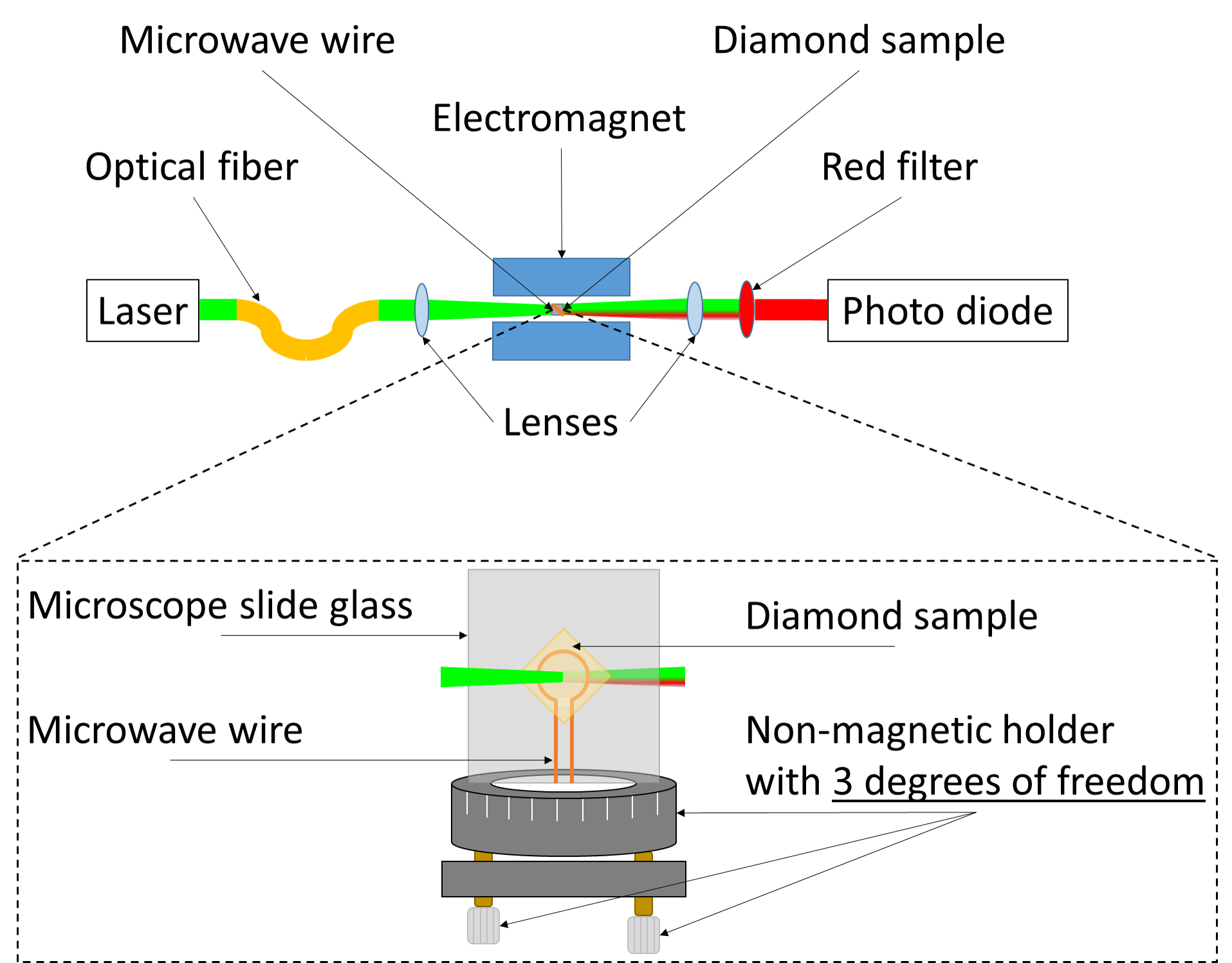}
  \end{center}
  \caption{Experimental setup. The top of the image shows the overall experimental scheme. The bottom of the image shows a zoomed-in, detailed scheme of the sample holder and the microwave wire.}
  \label{fig_exp_setup}
\end{figure}

Fig.~\ref{fig_exp_setup} shows the experimental setup used for the ODMR measurements.
The experimental device consisted of a 3-axis adjustable nonmagnetic sample holder (custom-made by STANDA) and two focusing aspheric condenser lenses for delivering (Thorlabs ACL12708U) and collecting (Thorlabs ACL25416U) the light to and from the diamond sample.

The nonmagnetic sample holder was attached to an aluminum rail, which was placed inside an electromagnet made of two iron poles $19$ cm in diameter with a length of $13$ cm each and separated by a $5.5$ cm air gap. Based on numerical modeling, we estimated that the magnetic field created by this electromagnet had an inhomogeneity of about $2\cdot10^{-3}$~G at 1000~G in a volume of about $2\times2\times0.1\text{ mm}^3$. The step size of the experimental magnetic field was determined by limitations of the power supply that was used to power the electromagnet (Agilent Technologies N5770A); it was about 0.9~G. This step size in combination with the available magnetic / MW field ranges resulted in about 1220 data points over the 0 to 1100~G magnetic-field range (different ODMR spectra for each magnetic field value).

The diamond sample was glued (using an adhesive Electron Microscopy Sciences CRYSTALBOND 509) to a custom-made microwave wire (an ``omega"-shaped sputtered copper trace, about 1~mm in diameter) etched on top of a microscope glass slide that was attached to the sample holder.
The green light was generated by a laser ($532$~nm, Coherent Verdi, Nd:YAG) and delivered to the experimental device using a single-mode optical fiber (Thorlabs 460HP). The laser power density that was delivered to the sample was approximately 3~kW/cm$^2$, as determined by the laser spot size on the sample ($\approx\! 100~\rm\upmu m$) and the laser power ($\approx\! 200$~mW) delivered to the experimental device.

The collected red fluorescence was filtered using a red longpass filter (Thorlabs FEL0600) and focused onto an amplified photodiode (Thorlabs PDA36A-EC).
The signals were recorded and averaged on a digital oscilloscope (Yokogawa DL6154).
A microwave generator (SRS SG386) in combination with a power amplifier (Minicircuits ZVE-3W-83+) was used to generate the necessary microwaves (MW) over a frequency range of 2.5--6.0 GHz  for the $m_S = 0\longrightarrow m_S=+1$ NV ground-state transition.

\section{Results and discussion}

\begin{figure*}
    \includegraphics[width=0.99\textwidth]{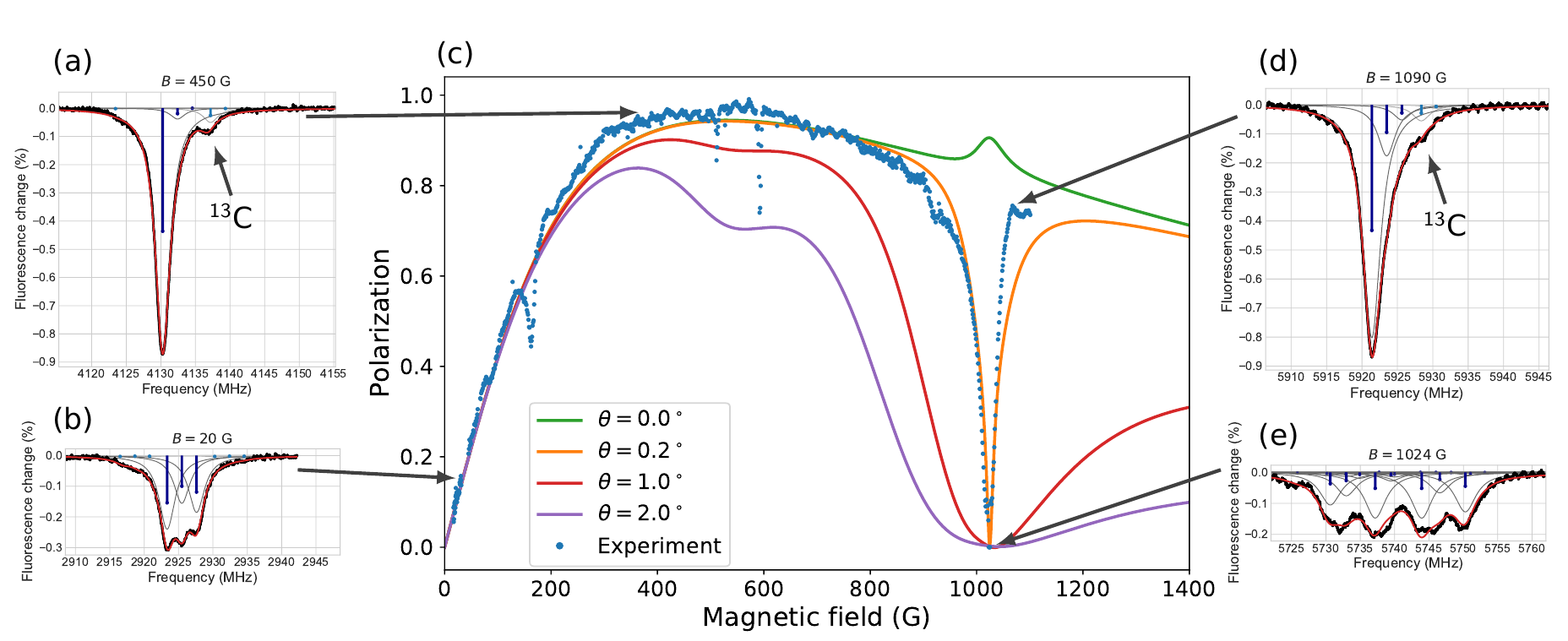}
  \caption{(a), (b), (d), (e): ODMR signals at individual magnetic field values. (c): Experimental (blue dots, each representing one ODMR measurement) and theoretical (solid curves) nuclear spin polarization as a function of magnetic field.
A more significant change depending on the angle can be seen at the GSLAC region rather than the ESLAC.
Pumping rate for theoretical calculations was $\Gamma_p = 5~\mathrm{MHz}$. The experimental ODMR curve fit determined that the magnetic field angle $\theta = 0.2^\circ$, which is in good agreement with the calculated curve at magnetic field angle $\theta = 0.2^\circ$.}
  \label{fig_pol_full}
\end{figure*}

The nuclear spin polarization was determined from the ODMR signals at every magnetic field value [Fig. \ref{fig_pol_full}(a),(b),(d),(e)], by fitting the experimental signals (black dots) with a modelled ODMR curve (red curve), which is a sum of Lorentzian curves whose central frequencies are located at the hyperfine transition frequencies $\nu_i$. 
The resonance amplitudes $t_i$ of the OMDR signal can be expressed as a product of transition probabilities $p_i$ and relative populations $N_i(m_I)$ of the hyperfine basis states $m_I$: $t_i = p_i N_i(m_I)$.

\begin{figure}[tb] 
    \includegraphics[width=0.45\textwidth]{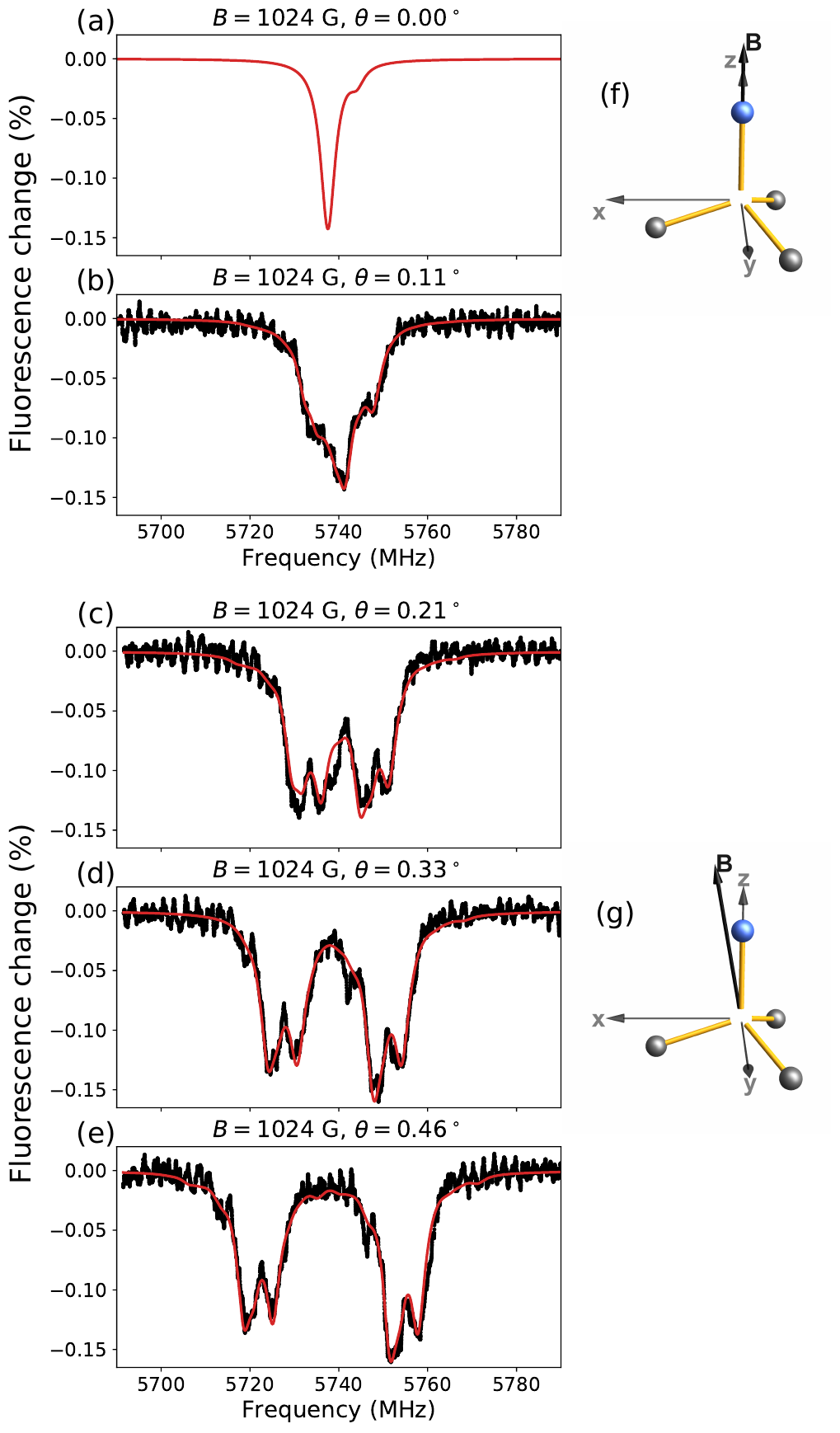}
  \caption{Left: ODMR signals at the GSLAC (1024~G) for different angles between the magnetic field and the NV axis; (a) shows the theoretical ODMR signal, whereas (b), (c), (d), and (e) show experimental (black dots) and fitted (red curve) ODMR signals. Right: representations of the external magnetic field direction with respect to the NV center's axis (f) for no transverse magnetic field or (g) a visually exaggerated transverse magnetic field. Plot (a) corresponds to the geometry shown in (f); plots (c), (d) and (e) correspond to the geometry shown in (g).}
  \label{fig_odmr_signals_theta_off}
\end{figure}

The hyperfine-level transition frequencies $\nu_i$ and transition probabilities $p_i$ were calculated from the eigenvalues and eigenfunctions of the ground-state Hamiltonian, Eq.~\eqref{eq_ham_ground}, taking into account the selection rules for MW transitions $\Delta m_S = \pm 1$ and $\Delta m_I = 0$. This approach allowed us to calculate the shape of the ODMR signals also at the GSLAC (Fig.~\ref{fig_odmr_signals_theta_off}), where the hyperfine levels are heavily mixed~\cite{Auzinsh2019}. The transition amplitudes between individual hyperfine components were modified by a fitting parameter $N_i(m_I)$, which corresponds to the relative population of the hyperfine basis state $m_I$. 

By fitting the experimental ODMR signal with this curve, we were able to determine the following parameters: individual Lorentzian resonance width (equal for all components), nuclear spin polarization, and magnetic field angle (magnetic field misalignment).

The experimental nuclear spin polarization of $^{14}$N was determined using the fitted relative populations of the hyperfine basis states:
\begin{equation}
P_{exp} = \frac{\sum_i m_I N_i(m_I)}{I\sum_i N_i(m_I)},\label{eq_exp_polarization}
\end{equation}
where the summation was performed over all of the fitted resonances of the experimental ODMR signal, $m_I$ is the nuclear-spin projection of the basis states and $I$ is the nuclear spin.
For magnetic field values far from the GSLAC, three hyperfine transitions were observed, so that each of the basis states of $m_I$ corresponds to a single transition. If the ground-state $\vert m_S=0\rangle$ and $\vert m_s=-1\rangle$ states undergo mixing, there are more than three allowed hyperfine transitions [Fig.~\ref{fig_pol_full}(e)], but from the hyperfine-interaction Hamiltonian, we can determine which base states $\vert m_S, m_I\rangle$ are involved. 
The sum of resonance amplitudes of each of the basis states corresponds to the relative population of the basis state, which allowed us to compare directly the experimentally determined nuclear spin polarization Eq.~\eqref{eq_exp_polarization} with the theoretically calculated nuclear spin polarization Eq.~\eqref{eq_theor_polarization}.
 
At a magnetic field value of around 20~G [Fig.~\ref{fig_pol_full}(b)] the amplitudes of three distinct hyperfine-transition components are almost equal, indicating a low $^{14}\mathrm{N}$ nuclear spin polarization. At a magnetic field close to the ESLAC [Fig.~\ref{fig_pol_full}(a)] the contrast of the ODMR signal is around three times larger due to nuclear spin polarization, and there is one dominant resonance transition. The population accumulation in this one resonance transition corresponds to a high $^{14}\mathrm{N}$ nuclear spin polarization of $96\pm 2\%$.

The diamond crystal consists mostly of $^{12}\mathrm{C}$, an isotope with zero nuclear spin, but $1.1\%$ of the carbon atoms belong to the $^{13}\mathrm{C}$ isotope. As the nuclear spin of $^{13}\mathrm{C}$ is $I_\text{C13} = 1/2$, it interacts with the electronic spin of the NV center, leading to an extra feature next to the main peak in the ODMR signal [see Fig. \ref{fig_pol_full}(a)]. The $^{13}\mathrm{C}$ interaction strength depends on its position in regard to the NV center \cite{Gali2008, Gali2009, Smeltzer2011, Nizovtsev2014}. Owing to the symmetry of the diamond crystal, the nearest $^{13}\mathrm{C}$ nucleus in the crystal lattice can have 3 equivalent positions and its hyperfine interaction strength is $A_\text{hfs} = 130~\mathrm{MHz}$. Although the interaction strength of these $^{13}\mathrm{C}$ nuclei allows these transitions to be resolved, the feature is too small to be observed in the experiment. 
The next closest families of possible $^{13}\mathrm{C}$ lattice sites have 3 and 6 equivalent positions with hyperfine interaction strengths of 12.8 and 13.8~MHz~\cite{Dreau2012, Nizovtsev2014}, respectively. The feature that comes from these $^{13}\mathrm{C}$ nuclei is clearly seen in the ODMR signal to the right of the main peak [see Fig.~\ref{fig_pol_full}(a)].
The hyperfine interaction of other $^{13}\mathrm{C}$ nuclei that are located at lattice sites further from the NV center is smaller than 10~MHz. The transition peaks from these sites are not resolved in our ODMR signals.

 The nuclear spin polarization was measured in a magnetic field range from 0 to 1100 G, which includes both the ESLAC and the GSLAC field regions [Fig.~\ref{fig_pol_full}(c)].
\begin{figure}[tb] 
    \includegraphics[width=0.45\textwidth]{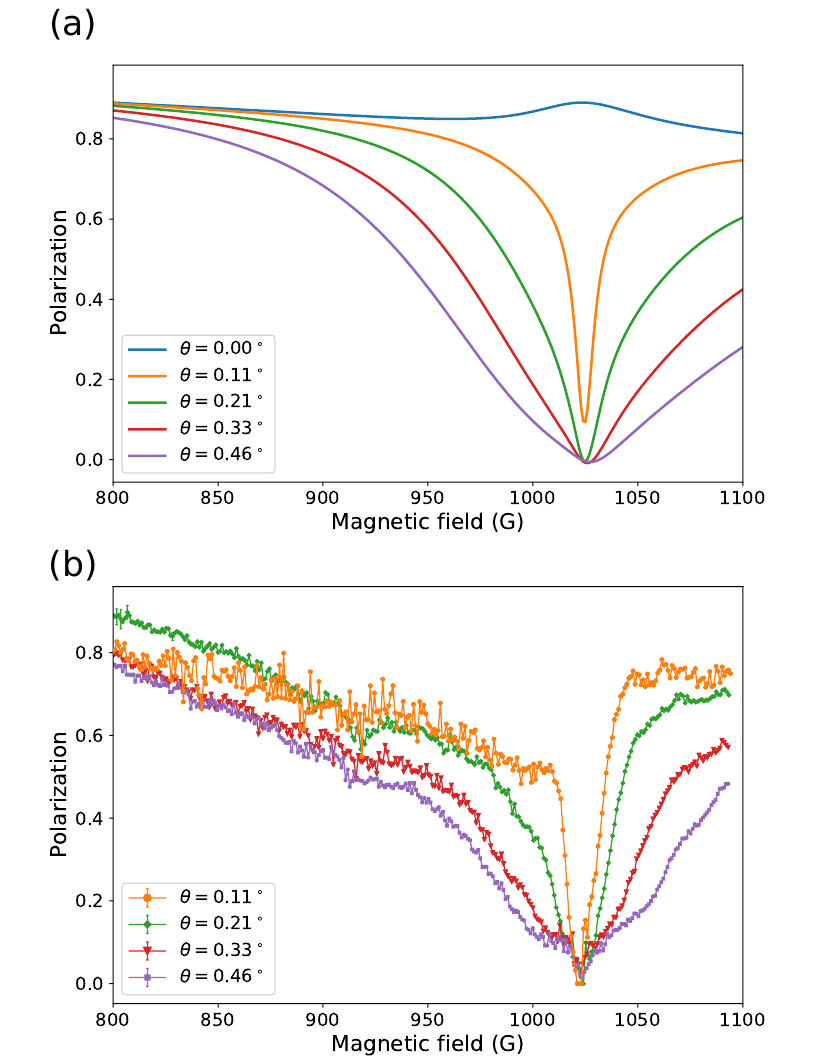}
  \caption{(a) Theoretical nuclear spin polarization around the GSLAC for different magnetic field angles $\theta$. The pumping rate for the numerical calculations was $\Gamma_p = 5~\mathrm{MHz}$. (b) Experimental nuclear spin polarization at the GSLAC for different magnetic field angles $\theta$.}
  \label{fig_pol_theta}
\end{figure}
At 512~G there is a drop in polarization due to the resonant interaction of the NV center with substitutional nitrogen in the lattice of the diamond crystal and cross-relaxation between the NV center and substitutional nitrogen~\cite{Hall2016, Negyedi2017,Lazda2020}.
At around 590 G the ground-state transition energy of the NV centers whose principal axis is parallel to the \textbf{B} direction matches the ground-state transition energy of the NV centers of the other three possible NV-axis orientations in the diamond crystal. Due to C$_{\text{3v}}$ symmetry, these directions are equivalent, and the NV centers' principal axis and magnetic field direction \textbf{B} form an $109.47^\circ$ angle. The energy matching and cross-relaxation at this magnetic field value contribute to the drop in the nuclear spin polarization~\cite{Anishchik2016}.
The decrease in polarization around 200~G occurs due to cross-relaxation between the NV center and other unknown paramagnetic defect centers in the crystal~\cite{VanOort1989, Anishchik2016, Anishchik2017}.

Our analysis shows that even a very small angle between the magnetic field direction and the NV center axis strongly influences the nuclear spin polarization near the GSLAC. This happens because the transverse magnetic field introduces additional level mixing, which strongly modifies the polarization process.
The solid curves in Fig.~\ref{fig_pol_full}(c) are calculated for different magnetic field angles. It can be seen that the GSLAC magnetic-field region is much more sensitive to magnetic field misalignment: for an angle as small as  $0.2^\circ$, the nuclear spin polarization is destroyed at the GSLAC while at the ESLAC magnetic field region only a small dip occurs for an angle as large as $1^\circ$.
Fig.~\ref{fig_pol_theta}(a) shows a theoretical calculation of the nuclear spin polarization for magnetic field angles ranging from $\theta =0^\circ$ to $\theta = 0.46^\circ$ near the GSLAC. While polarization near the ESLAC (512 G) [Fig.~\ref{fig_pol_full}(c)] is not affected by such small angles because of the much stronger hyperfine interaction there~\cite{Poggiali2017, Smeltzer2009, Felton2009}, the polarization at magnetic field values near the GSLAC (1024 G) is $\sim\! 85\%$ for $\theta=0^\circ$; it decreases significantly for $\theta=0.11^\circ$ and approaches almost $0\%$ for $\theta = 0.21^\circ$. 
The experimental measurements of nuclear spin polarization show the same dependence on the magnetic field angle [Fig.~\ref{fig_pol_theta}(b)] as suggested by the theoretical calculations.

To align the principal axis of the NV center along the external magnetic field direction, we used the ODMR signals acquired at the GSLAC (Fig.~\ref{fig_odmr_signals_theta_off}). Our fitting procedure allowed us to determine the angle of the magnetic field with a precision of $0.02^\circ$. Perfect alignment [see visualization in Fig.~\ref{fig_odmr_signals_theta_off}(f)] is achieved when the ODMR signal consists of one group of peaks. Fig.~\ref{fig_odmr_signals_theta_off}(a) shows a theoretical simulation of an ODMR signal for a $0^\circ$ angle between the NV axis and the magnetic field vector. The nuclear spin polarization for this curve is $85\%$, as determined from the Lindblad calculation [blue line in Fig.~\ref{fig_pol_theta}(a)]. The Lorentzian width and other parameters are the same as for the fitted curves.
Fig.~\ref{fig_odmr_signals_theta_off}(b) shows the ODMR signal for the best experimentally achieved alignment with respect to the external magnetic field direction, which corresponds to $0.11^{\circ}$.
Increasing the angle between the NV axis and the external magnetic field direction [see visualization in Fig.~\ref{fig_odmr_signals_theta_off}(g)] causes the ODMR structure to transform into two groups of peaks, which is explained by the mixing of the $\vert m_S=0\rangle$ and $\vert m_S=-1\rangle$ energy levels, which are energetically close at the GSLAC point. At angles between $0.21^{\circ}$ and $0.46^{\circ}$ two distinct groups of peaks can be seen in the ODMR signal [Fig.~\ref{fig_odmr_signals_theta_off}(c)--(e)].

\begin{figure}[tb] 
    \includegraphics[width=0.45\textwidth]{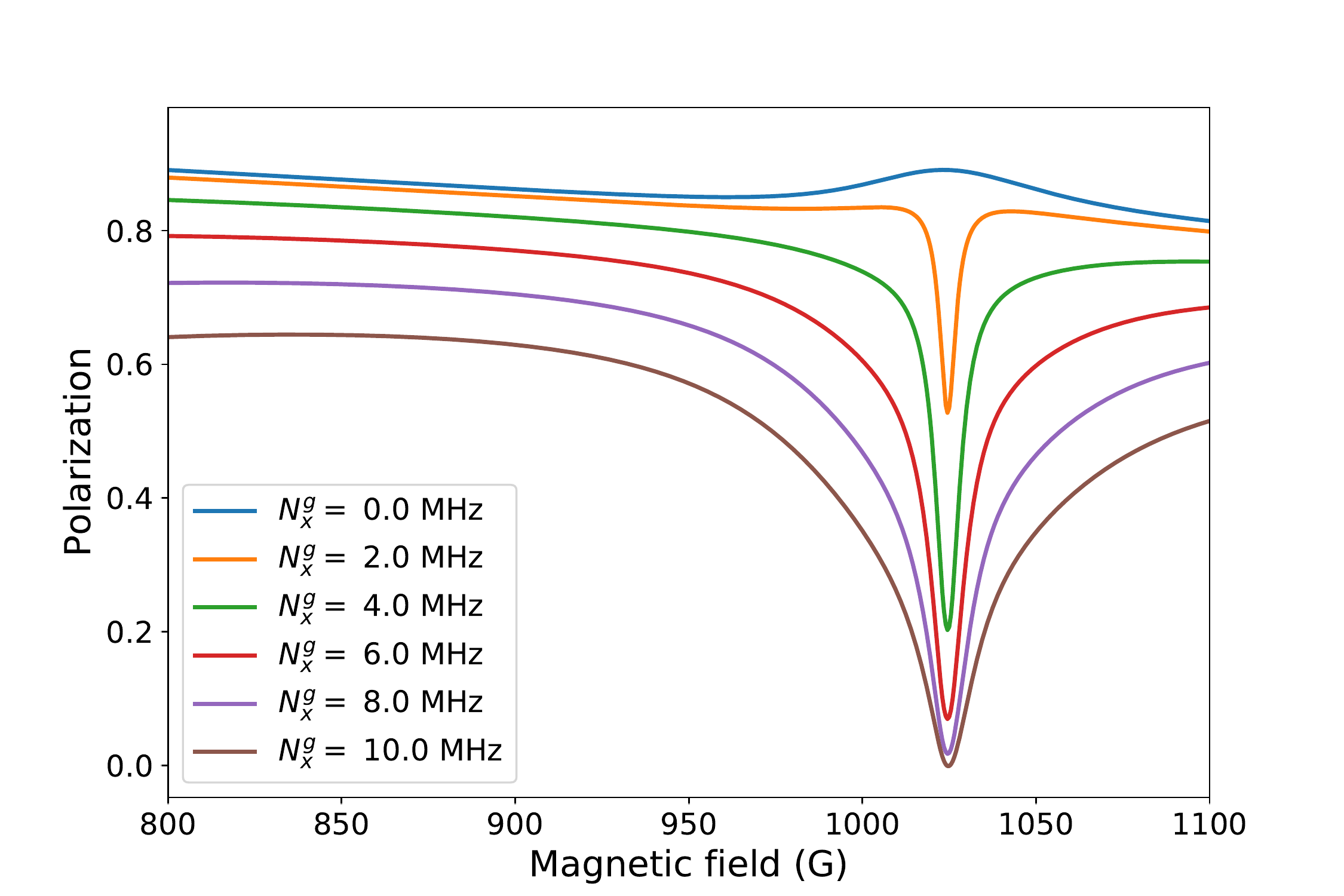}
  \caption{Theoretical nuclear spin polarization around the GSLAC for different transverse strain $N_x$ values. The pumping rate in the numerical calculations was $\Gamma_p = 5~\mathrm{MHz}$. The excited-state strain was set to $N^e_x=10N^g_x$.}
  \label{fig_pol_theor_strain}
\end{figure}

The spin-strain interaction \cite{Doherty2012, Udvarhelyi2018, Barfuss2019, Kehayias2019} can cause a similar effect to the transverse magnetic field.
A transverse strain component $N_x$ leads to coupling between the $m_s= 0$ and $m_s=\pm 1$ levels [see Eq.~\eqref{eq_ham_ground}]. At the GSLAC, where the energy difference between levels $m_s=0$ and $m_s = -1$ is small, the ground-state strain component $N^g_x$ gives a result similar to what would be expected from a transverse magnetic field (see Fig.~\ref{fig_pol_theor_strain}).
The influence of the ground-state strain $N^g_x$ was investigated by theoretical calculations. As the ground-state stain values are typically few megahertz, while the excited-state strain is a few tens of megahertz~\cite{Doherty2013, Fuchs2008}, the excited-state strain was set to $N^e_x=10N^g_x$ in the calculations.
The strain present in the crystal could be the reason that prevents achieving an experimental ODMR signal as Fig. \ref{fig_odmr_signals_theta_off}(a). Since strain affects the nuclear spin polarization, this also could be the reason for not achieving high polarization at the GSLAC.

\begin{figure}[t] 
    \includegraphics[width=0.45\textwidth]{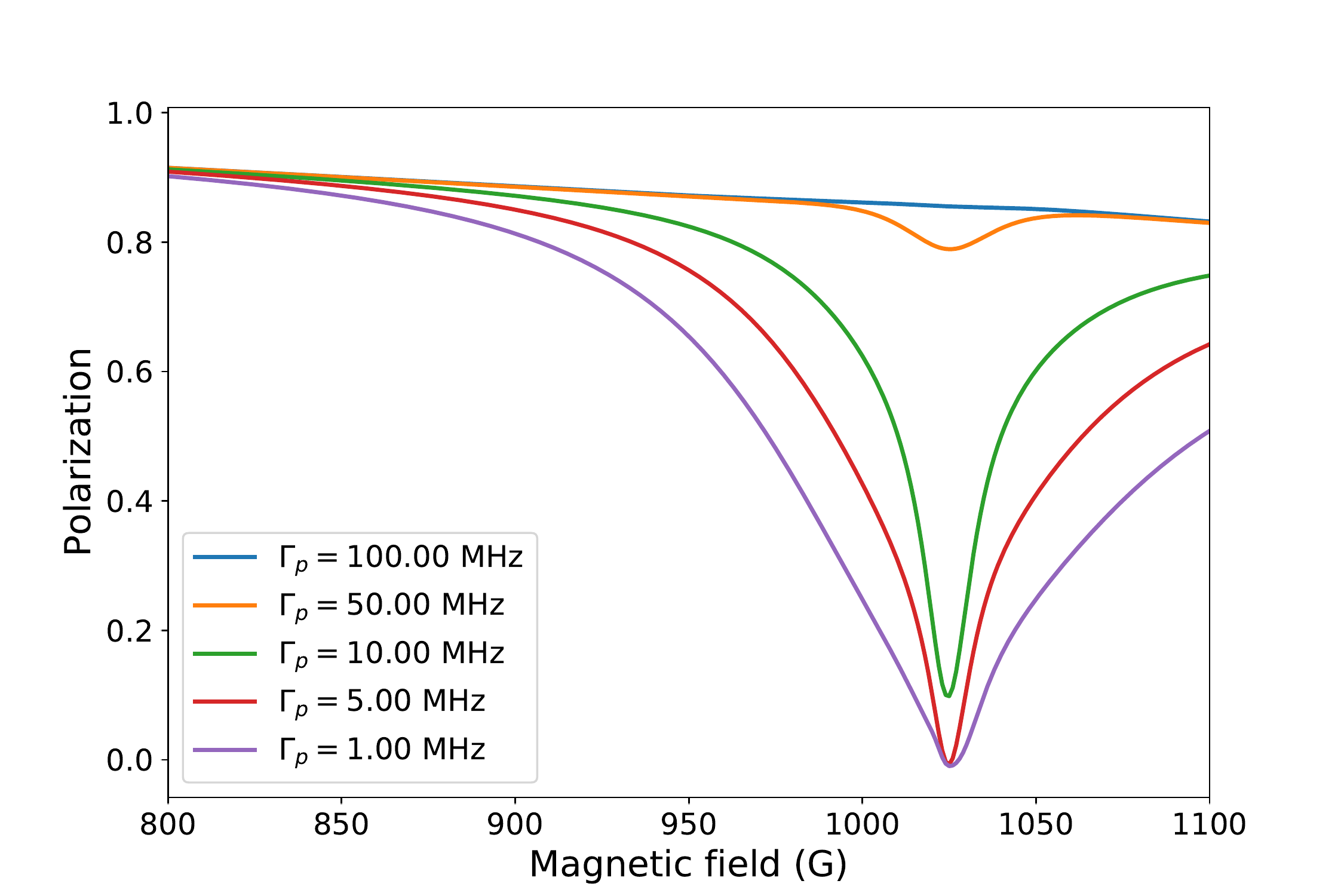}
  \caption{Theoretical nuclear spin polarization around the GSLAC at different pumping rate $\Gamma_p$ values for $\theta=0.2^\circ$.}
  \label{fig_pol_theor_power}
\end{figure}

Nuclear spin polarization at the GSLAC is also sensitive to the pumping laser power, which corresponds to the pumping rate $\Gamma_p$ in the model.
Fig. \ref{fig_pol_theor_power} shows the calculated nuclear spin polarization for different pumping rates for a magnetic field angle $\theta=0.2^\circ$.
Increasing the pumping rate would increase the polarization, whereas increasing the strain tends to decrease the polarization (see Fig.~\ref{fig_pol_theor_strain}). Therefore, the two effects can be decoupled in the theoretical fitting procedure to some extent.

\section{Conclusions}

As was discussed in the Introduction, dynamic nuclear spin polarization attracts attention not only as an academically interesting effect that takes place in NV centers in diamond crystals in the vicinity of the ESLAC and GSLAC, but it also plays an important role in different applications with potential for quantum technologies~\cite{Aharonovich2011} and as a way to strongly enhance the sensitivity of NMR methods.
In this study we used the ODMR technique to measure the dynamic nuclear spin polarization of nitrogen, which is part of the NV center, over a broad range of magnetic field values that included both the ESLAC at 512~G and GSLAC at 1024~G. We measured the amplitudes of ODMR signal peaks and compared the results of these measurements with a theoretical model to determine the nuclear spin polarization.

These measurements show that with optimal magnetic field alignment a very large ($>\!80\%$, up to $96\pm 2\%$) nuclear spin polarization of nitrogen can be achieved over a very broad range of magnetic field starting from around 400~G up to magnetic field values that substantially exceed the GSLAC at 1024~G. 
It appears that the DNP of nitrogen is very sensitive to a number of factors. Even a slight misalignment between the applied magnetic field direction and the axis of the NV center by less than $0.1^\circ$ can almost completely destroy the nuclear spin polarization in the vicinity of the GSLAC (Fig.~\ref{fig_pol_theta}). Moreover, in the bulk sample with many NV centers, which, perhaps due to strain in the crystal, can have slightly different orientations of their axis, such perfect alignment between the magnetic field and the axis of the NV centers was not possible to achieve with the diamond sample that we used (Fig.~\ref{fig_pol_theor_strain}).

The results obtained in this paper shed light on the less-studied nuclear spin polarization, in particular in the GSLAC region. The results were obtained using a simple method that permits the evaluation of nuclear spin polarization in a rather complex level-crossing situation. 
The results should contribute to a better understanding of DNP processes in NV centers, which can make NV applications more efficient.

\section{Acknowledgements}
We gratefully acknowledge the financial support from the Base/Performance Funding Project Nr. ZD2010/AZ27, AAP2015/B013 of the University of Latvia. A.~Berzins acknowledges support from the PostDoc Latvia Project Nr. 1.1.1.2/VIAA/1/16/024 "Two-way research of thin-films and NV centres in diamond crystal". The authors are thankful to Dr. I.~Fescenko for useful discussions and comments.

\bibliographystyle{apsrev}
\bibliography{bibl-nuclear}

\end{document}